\documentclass[aps,prl,twocolumn,showpacs,groupedaddress,amsmath,amssymb]{revtex4}

\usepackage{graphicx}
\usepackage{color}
\usepackage{comment}
\usepackage{bm}
\usepackage{latexsym}

\usepackage[symbol]{footmisc}
\usepackage{hyperref}
 \usepackage{braket}

\usepackage[T1]{fontenc}
\usepackage[latin9]{inputenc}
\usepackage{amsmath}
\usepackage{esint}

\begin{document}

\title{Statistical Description of Transport in Multimode  Fibers with Mode-Dependent Loss}
\author{P. Chiarawongse$^{1}$\footnote{These authors contributed equaly in the theoretical analysis},  H. Li$^{1,*}$,
W. Xiong$^2$\footnote{This author performed the experiments}, C. W. Hsu$^{2,*}$, H. Cao$^2$\footnote{These authors 
supervised the project}, T. Kottos$^{1,\ddagger}$}
\address{$^1$Physics Department, Wesleyan University, Middletown CT-06459, USA}
\address{$^2$Department of Applied Physics, Yale University, New Haven CT-06520, USA}

\begin{abstract}
We analyze coherent wave transport in a new physical setting associated with multimode wave systems where reflection is completely 
suppressed and mode-dependent losses together with mode-mixing are dictating the wave propagation. An additional physical 
constraint is the fact that in realistic circumstances the access to the scattering (or transmission) matrix is incomplete. We have addressed
all these challenges by providing a statistical description of wave transport which fuses together a free probability theory approach 
with a Filtered Random Matrix ensemble. Our theoretical predictions have been tested successfully against experimental data of 
light transport in multimode fibers. 
\end{abstract}
\maketitle

Random Matrix Theory (RMT) has been successfully applied over the years in a variety of physics areas ranging from nuclear and atomic
physics to mesoscopic physics of disordered and chaotic systems \cite{ABF10,S99,B97,A00,EM08,S15}. Its applicability relies on the assumption 
that in complex systems the underlying wave interference impose universal {\it statistical} rules which govern their transport characteristics. 
Along these lines of thinking, random matrix models allowed us to uncover some of the most fundamental properties of disordered/chaotic 
systems, including the structure and statistical properties of their eigenstates \cite{FM94,BKS10} and eigenvalues \cite{AS86,CIM91}, the 
conductance \cite{BM94,I86,CGIMZ94}, the resonance widths and delay times \cite{FS97}, etc. It turned out that many of the universal features 
of transport are directly connected with the various symmetries (time-reversal, chiral, etc) that a specific complex system satisfies \cite{Z96}. 
In all these studies, nevertheless, it was always assumed that the scattering process does not involve any additional constrains and has both 
a backward (reflection) and a forward (transmission) component. 

Recently, the interest in wave transport has extended to new physical settings with practical relevance, namely, a class of complex multimode 
systems where reflection processes are absent \cite{HK13,HK11}. Obviously, the zero reflectivity condition, imposes new constraints to the 
wave scattering process, thus constituting the previous RMT predictions void. These type of transport problems have emerged naturally in the 
framework of multi-mode (or coupled multi-core) fiber optics. In these systems, fiber imperfections (core ellipticity and eccentricity) and external 
perturbations (index fluctuations and fiber bending) cause coupling and interference between propagating signals in different spatial modes 
and orthogonal polarizations. At the same time, the effect of {\it mode-dependent} loss (or gain due to optical amplifiers) in wave propagation 
is another important feature whose ramifications are not yet completely understood \cite{HK13,HK11,Cao1,Cao2}. In the framework of multimode 
fibers (MMFs), for example, it leads to fundamental limitations in their performance since extremely  high mode-dependnet losses (MDL) can 
reduce the number of propagating modes and thus the information capacity of MMFs. It is, therefore, imperative to develop statistical theories 
that take into consideration the modal and polarization mixing and mode-dependent losses (MDL)  and provide a quantitative description of 
light transport in realistic MMFs (and other multimode systems that demonstrate similar challenges).

Here, we develop a statistical theory of light transport in MMFs, where both MDL and modal and polarization mixing are considered. Our theory 
is accounting for the fact that in experiments, the degree of modal and polarization control is limited. To this end, we have combined free probability 
theory and the Filtered Random Matrix (FRM) ensemble and took into consideration the finite length of the MMFs. Unlike the telecom fibers, 
which are typically very long, finite length MMFs are common in medical applications (endoscopy), sensing, local-area networks and data-center 
interconnects etc. As an example, we have implemented our theoretical formalism in order to derive two important  statistical measures: (a) the 
distribution of transmission eigenvalues for polarization maintenance (and/or conversion); and (b) the absorbance distribution of a monochromatic 
light propagating in a MMF with MDL and strong mode and polarization coupling. Our theoretical results have been validated via direct comparison 
with experimental measurements using MMFs.

We consider a MMF supporting $N$ propagating linearly polarized (LP) modes, with each LP mode being two-fold degenerate corresponding to 
the horizontal (H) and vertical (V) polarizations. We model the fiber of interest as consisting of a concatenation of $K$ independent and statistically 
identical segments~\cite{HK11}, with the linear propagation through the MMF described by a $2N\times2N$ transmission matrix $t^{(K)}$,
\begin{align}
t^{(K)}= \begin{pmatrix}t_{\rm HH}^{(K)} & t_{\rm HV}^{(K)}\\t_{\rm VH}^{(K)} & t_{\rm VV}^{(K)}\end{pmatrix}=& \Pi^{K}v_{0},\quad\Pi^{K}
=m_{K}\cdots m_{2}m_{1},
\label{eq: Ttot}
\end{align}
where the elements of the $N\times N$ block matrices $t_{\rm HH}^{(K)}$ ($t_{\rm VH}^{(K)}$) are the transmission amplitudes into the H (V) 
polarization when the incident light is H-polarized. Each segment is modeled via a matrix $m_{k}=v_{k}\Lambda$, where $v_{k}$ is a $2N\times2N$ 
unitary matrix describing the polarization and mode mixing in the segment, and $\Lambda = {\rm diag}(\Lambda^{\rm H}, \Lambda^{\rm V})$ is a 
diagonal matrix describing the free propagation and attenuation in the absence of such mixing. We consider MMFs in the strongly 
mixed regime where every mode is coupled to every other mode in one segment, with $v_{k}$ and $v_0$ being random unitary matrices drawn 
from the circular unitary ensemble (CUE)~\cite{B97}. We assume that the two polarizations have the same propagation constants and loss, 
so that $\Lambda_{n,l}^{\rm H}=\Lambda_{n,l}^{\rm V}=e^{i\beta_n}\delta_{nl}$. The higher-order LP modes take longer paths and impinge on the 
core-cladding interface at steeper angles, so they typically experience more attenuation than the lower-order modes; we model 
such mode-dependent losses as ${\cal I}m\left(\beta_n\right)=ns/(2N)$ with $n=1,\cdots, N$, characterized by the  coefficient $s$ ($s>0$ 
for loss). The real parts of $\beta_n$ describe the mode-dependent propagation phase shifts and are not important in the context 
of this paper as they can be absorbed into $v_k$.

In actual experimental circumstances, the preparation and measurement of a waveform in all modes is technically challenging. In this respect,  one needs to analyze portions of the total transmission matrix $t_{P_{\rm out},P_{\rm in}}^{(K)}=P_{\rm out}t^{(K)}P_{\rm in}$ where $P_{\rm in}$ 
and $P_{\rm out}$ are projections to the controlled incoming and outgoing modal subspace. Specifically, given an incident wavefront $|\psi \rangle$ 
which belongs to the $P_{\rm in}$-subspace, the measured transmittance in the $P_{\rm out}$-subspace (summed over the spatial/polarization 
modes) after propagating through the MMF is $\langle \psi| \left(t_{P_{\rm out},P_{\rm in}}^{(K)}\right)^{\dagger}t_{P_{\rm out},P_{\rm in}}^{(K)}|\psi 
\rangle$. It is therefore obvious that the eigenvalues of the matrix $\left(t_{P_{\rm out},P_{\rm in}}^{(K)}\right)^{\dagger}t_{P_{\rm out},P_{\rm in}}^{
(K)}$ dictates the transport properties of such MMFs. For example, the extremal eigenvalues (and corresponding eigenvectors) are associated with 
the maximal and minimal transmittances achieved in such set-ups and can be used in order to design waveform schemes with extreme transport 
characteristics. Along these lines of reasoning, of particular interest is the eigenvalue statistics ${\cal P}_{\rm HH}^{(K)}(\tau)$ associated with the 
matrix $\left(t_{\rm HH}^{(K)}\right)^{\dagger}t_{\rm HH}^{(K)}$. In this case the preparation (associated with $P_{\rm in}$) and measurement 
(associated with $P_{\rm out}$) subspaces correspond to the set of modes with horizontal (H) polarization. The maximum eigenvalue $\tau$ (and 
the associated eigenvector) indicate the optimal polarization retention that can be achieved when light propagates in the system.  

Another interesting statistics is ${\cal P}_{\rm H}^{(K)}(\tau)$ associated with the matrix $T_H\equiv \left(t_{\rm HH}^{(K)}\right)^{\dagger}t_{\rm 
HH}^{(K)}+\left(t_{\rm VH}^{(K)}\right)^{\dagger} t_{\rm VH}^{(K)}$. In this case $P_{\rm in}$ corresponds to the subspace of holizontaly polarized 
modes while $P_{\rm out}$ is the identity matrix i.e. the whole modal space including both polarizations. The eigenvalues of $T_H$ provide 
information about the total transmissivity summed over the two polarization states at the output, given a H-polarized incident light. The complementary 
matrix $A_H\equiv 1- T_H$ provides information about the amount of absorption during propagation inside the MMF. 

The evaluation of the transmission eigenvalue distribution for any portion of the transmission matrix involves two steps. The first one is 
generic: it requires the calculation of the probability distribution ${\cal P}^{(K)}(\tau)$ of the eigenvalues of $\left(t^{(K)}\right)^{\dagger}
t^{(K)}=v_0^{\dagger}m_1^{\dagger}\cdots m_K^{\dagger} m_K\cdots m_1 v_0$ and the associated Green's function $G^{(K)}\left(z\right)
\equiv\int d\lambda\frac{{\cal P}^{(K)} \left(\lambda\right)}{z-\lambda}$ (also called the resolvent or the Stieltjes transform). We take advantage 
of the multiplicative structure of the transmission matrix, using free probability theory~\cite{TV04,Voiculescu,Burda} which predicts the spectral properties 
of a product of random matrices from the spectral properties of its constituents. Based upon the probability distribution ${\cal P}^{(K=1)}
(\tau)={\cal P}_{\Lambda^{2}}\left(\tau\right)= 1/({s\tau})$ associated with the eigenvalues of $\left(t^{(1)}\right)^{\dagger}t^{(1)}$ for a single 
segment, we show in the supplement that
\begin{align}
\frac{1}{z}= & \left(\frac{zG^{(K)}}{zG^{(K)}-1}\right)^{K-1}\left(\frac{e^{szG^{(K)}}-e^{s}}{e^{szG^{(K)}}-1}\right)^{K}.
\label{eq: g_tK}
\end{align}
In the second step, we use $G^{(K)}$ to derive the Green's function $G_{P_{\rm in},P_{\rm out}}^{(K)}$ and the eigenvalue distribution 
${\cal P}_{P_{\rm in},P_{\rm out}}^{(K)}(\tau)$ associated with the projected transmission matrix $t_{P_{\rm out},P_{\rm in}}^{(K)}$. Following 
a Filtered Random Matrix (FRM) formalism~\cite{Goetschy}, we have that 
\begin{align}
G^{(K)}\left({n_{P_{\rm out}, P_{\rm in}}^2\over d_{P_{\rm out},P_{\rm in}}}\right)={d_{P_{\rm out},P_{\rm in}}\over n_{P_{\rm out},P_{\rm in}}}
\label{eq: FRM}
\end{align}
where $n_{P_{\rm out}, P_{\rm in}}$ and $d_{P_{\rm out},P_{\rm in}}$ are two auxiliary functions related to the filtering process 
$P_{\rm out},P_{\rm in}$ and the Green's function $G_{P_{\rm out},P_{\rm in}}^{(K)}$. 

For the specific case of $G_{\rm HH}^{(K)}$ (and the eigenvalue distribution ${\cal P}_{\rm HH}^{(K)}(\tau)$ of $\left(t_{\rm HH}^{(K)}
\right)^{\dagger}t_{\rm HH}^{(K)}$) we have that $n_{\rm HH}={1\over 2} \left(zG_{\rm HH}^{(K)}+1\right)$ and $d_{\rm HH}={1\over 4}
z\left(G_{\rm HH}^{(K)}\right)^2$. For this derivation one needs to consider that the input and output fraction of total mode space is half. 
Combining Eqs.~(\ref{eq: g_tK},\ref{eq: FRM}), we derive an implicit equation
\begin{align}
\frac{1}{z}= & \left(\frac{p_{\rm HH}}{p_{\rm HH}-1}\right)^{2}\left(\frac{p_{\rm HH}}{p_{\rm HH}-2}\right)^{K-1}\left(\frac{e^{\frac{s}{2}
p_{\rm HH}}-e^{s}}{e^{\frac{s}{2}p_{\rm HH}}-1}\right)^{K},
\label{eq: HH}
\end{align}
which we can solve to obtain $p_{\rm HH}\left(z\right)\equiv zG_{\rm HH}^{(K)}\left(z\right)+1$ and $G_{\rm HH}^{(K)}\left(z\right)$.
Then, the probability distribution ${\cal P}_{\rm HH}^{(K)}(\tau)$ is given by the inverse Stieltjes transform 
\begin{equation}
{\cal P}_{\rm HH}^{(K)}\left(\lambda\right)=  -\frac{1}
{\pi}\lim_{\epsilon\rightarrow0^{+}}\mathrm{Im}G_{\rm HH}^{(K)}\left(\lambda+i\epsilon\right).
\label{GfuncP}
\end{equation}
The above analysis also captures the effect of incomplete modal control, {\it e.g.} when only parts of the $N$ spatial modes are 
modulated or measured.

\begin{figure}[h]
\includegraphics[width=1\columnwidth,keepaspectratio,clip]{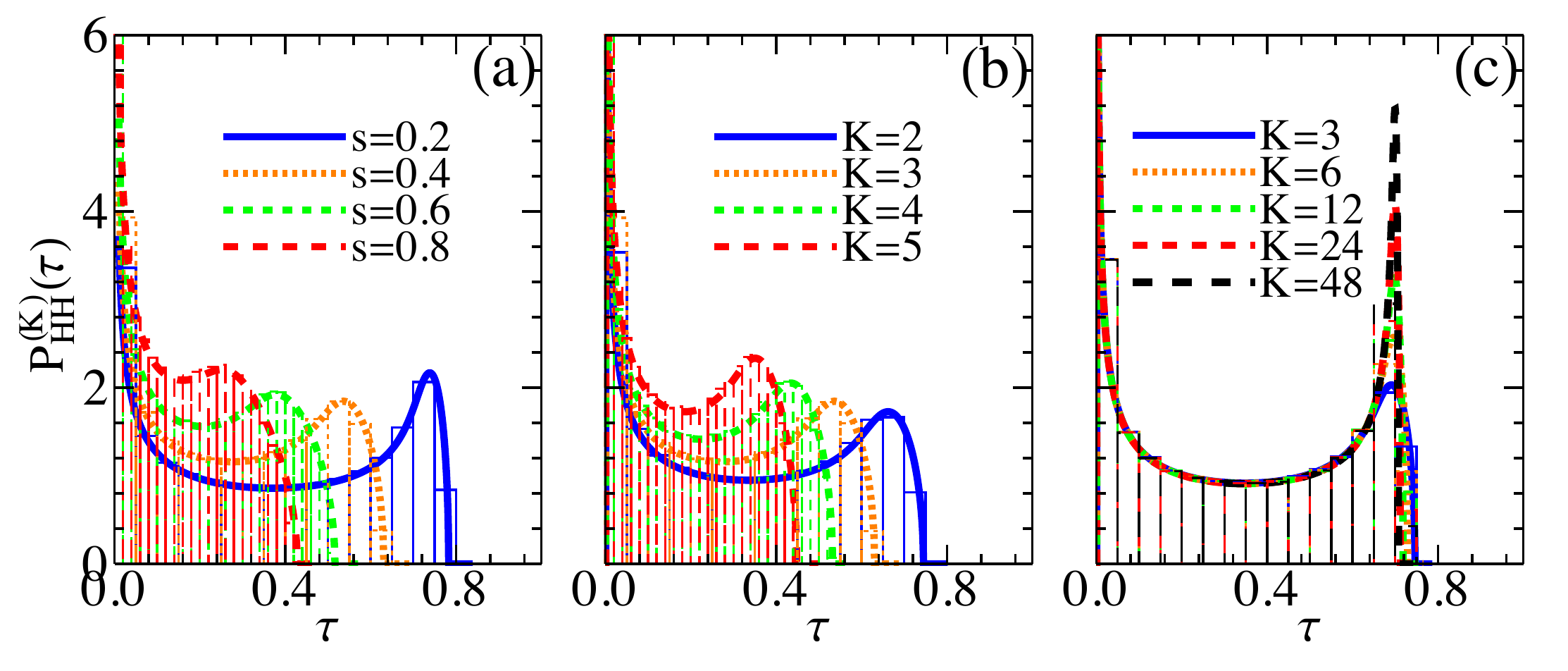}
\caption{ (color online) Eigenvalue distributions ${\cal P}_{\rm HH}^{(K)}(\tau)$ for the polarization-maintaining states in a multimode fiber.
(a) Dependence on the loss-per-segment parameter $s$ for a fixed number of segments $K=3$.
(b) Dependence on the number of segments $K$ for a fixed $s=0.4$.
(c) The long-fiber limit $K\rightarrow \infty$, $s\rightarrow 0$ while keeping the overall loss $\mu^{(K)}=0.7$ fixed.
Solid lines are analytic predictions from Eqs.~(\ref{eq: HH},\ref{GfuncP}), and histograms are from numerical simulations of the the 
concatenated MMF model with $N=25$ modes for (a-b) and $N=75$ for (c).
}
\label{fig1}
\end{figure}

Using Eqs.~(\ref{eq: HH},\ref{GfuncP}) we can derive analytical expressions for ${\cal P}_{\rm HH}^{(K)}(\tau)$ for various $s$ 
values and number of concatenated segments $K$, see Fig.~\ref{fig1}. We find that for increasing loss-per-segment 
parameter $s$, the deviations from a bimodal distribution ${\cal P}_{s=0}^{(\infty)}\left(\tau\right)$ become progressively stronger, see 
Fig.~\ref{fig1}a. The same is true for the case of increasing number of concatenated segments $K$ while keeping $s$ fixed, see 
Fig.~\ref{fig1}b. In both cases, the most dramatic changes occur at the upper edge of ${\cal P}_{\rm HH}^{(K)}(\tau)$ associated with 
the largest transmission eigenvalues. In the same figures, we also plot the histograms from numerical simulations of the concatenated 
MMF model with a finite number of modes. The agreement between the theoretical predictions and the numerical simulations is perfect. 

Furthermore, the first two moments of the eigenvalue distributions ${\cal P}^{(K)}(\tau)$ and ${\cal P}_{HH}^{(K)}(\tau)$ can be analytically 
derived and they acquire a simple closed form (see Supplement). We get that the mean $\mu^{(K)}$ and variance $\left(\sigma^{(K)}
\right)^{2}$ of ${\cal P}^{(K)}(\tau)$ are $\mu^{(K)}= \mu^{K}$ and $\left({\sigma^{(K)}}/{\mu^{(K)}}\right)^2=K{\sigma^{2}}/{\mu^{2}}$ 
respectively; these moments characterize the strength and the spread of the overall loss. Here $\mu=\left({1-e^{-s}}\right)/{s}$ and 
$\sigma^{2}=\frac{1-e^{-2s}}{2s}-\left(\frac{1-e^{-s}}{s}\right)^{2}$ are the mean and variance of ${\cal P}^{(K=1)}(\tau)$ for a single 
segment. Similarly, we also show in the Supplement that the mean $\mu_{\rm HH}^{(K)}$ and variance $\left(\sigma_{\rm HH}^{(K)}
\right)^{2}$ of the polarization-maintaining eigenvalue distribution ${\cal P}_{\rm HH}^{(K)}\left(\tau\right)$ are $\mu_{\rm HH}^{(K)}= 
\mu^{K}/2$ and $\left({\sigma_{\rm HH}^{(K)}}/{\mu_{\rm HH}^{(K)}}\right)^2=\frac{1}{2}\left[(K{\sigma^{2}}/{\mu^{2}})+1\right]$.

The explicit knowledge of the first two moments allow us to analyze the scenario of many concatenated fiber segments $K\to\infty$ 
with a loss-per-segement $s\to 0$, such that the mean $\mu^{(K)}\equiv \int {\cal P}^{(K)}(\tau)\tau d \tau$ is kept fixed, i.e. $\mu^{(K)}
={\cal C}$. We find, using the Bhatia-Davis inequality, that in this case the variance of the probability distribution ${\cal P}^{(K)}(\tau)$ 
goes to zero $\sigma^{(K\rightarrow\infty)}\to 0$. Consequently, the eigenvalue distribution ${\cal P}^{(\infty)}\left(\tau\right)\rightarrow
\delta\left({\cal C}-\tau\right)$ becomes a delta function and thus $G^{(\infty)}(z)={1\over z-{\cal C}}$. Then, using Eq. (\ref{eq: FRM}) 
for the specific case of $G_{\rm HH}^{(\infty)}$, we get $G_{\rm HH}^{(\infty)}(z)= {1\over \sqrt{z(z-{\cal C})}}$. Consequently we find that 
${\cal P}_{\rm HH}^{(\infty)}\left(\tau\right) \to \frac{1}{\pi\sqrt{\tau\left({\cal C}-\tau\right)}}$ reduces to a bimodal distribution with 
confined support $\tau\in\left(0,{\cal C}\right)$. In this case the information about the number of concatenation segments $K$ and
the loss-per-segment $s$ is ``hidden'' in the upper bound of the transmittance support $C$. In the  limiting case of zero losses $s=0$, 
it is easy to show, that the eigenvalue distribution ${\cal P}_{\rm HH}\left(\tau\right)$ reduces to a bimodal distribution ${\cal P}_{s=0}
\left(\tau\right)=\frac{1}{\pi\sqrt{\tau\left(1-\tau\right)}}$~\cite{XHBELCC17}. 

In the absence of loss $s=0$, it is easy to show, that the eigenvalue distribution ${\cal P}_{\rm HH}^{(\infty)}\left(\tau\right)$ reduces 
to a bimodal distribution ${\cal P}_{s=0}^{(\infty)}\left(\tau\right)=\frac{1}{\pi\sqrt{\tau\left(1-\tau \right)}}$~\cite{XHBELCC17}. It is tempting, 
at this point, to establish an analogy between the $s=0$ and $s\ne0$ cases ($K\rightarrow \infty$). In both cases there are essentially 
only two groups of propagating channels -- open channels associated with $\tau$-values close to 1 or ${\cal C}$, and closed channels 
with $\tau$-values in the neighborhood of zero. One then can understand the results for $K\rightarrow \infty$, $s\rightarrow 0$  in the 
following way: when the MMF is long enough (large $K$) such that complete mode and polarization mixing happens many times across 
the fiber, the mixing equalizes the mode dependence and turns MDL into mode-independent losses with the transmittance of the open 
channels being renormalized to ${\cal C}$. These analytic predictions are nicely confirmed by numerical simulations of the concatenated 
MMF model, as shown in Fig.~\ref{fig1}c.

We finally note that because of the strong mode and polarization mixing, this analysis applies equally to ${\cal P}_{\rm VH}^{(K)}(\tau)$ 
or other quarters of the transmission matrix. We stress that the calculation strategy that we have used here is not bounded by the specific
choice of MDL (constant increase) and can be easily generalized to any type of MDL distribution. Moreover, the same scheme can be 
utilized for the case of mode-dependent gain.

Using the same approach as above, we can also evaluate the eigenvalue distribution ${\cal P}_{\rm H}^{(K)}(\tau)$ of matrix $T_H\equiv 
\left(t_{\rm HH}^{(K)}\right)^{\dagger}t_{\rm HH}^{(K)}+\left(t_{\rm VH}^{(K)}\right)^{\dagger} t_{\rm VH}^{(K)}$ and the associated mean 
and variance. These results are given by Eqs.~(S10,S11) in the Supplement. These eigenvalues provide information about the total 
transmissivity summed over the two polarization states at the output, given a H-polarized incident light. Alternatively, one can consider 
the complementary matrix $A_H=1-T_H$ whose eigenvalues $\alpha\equiv1-\tau$ provide the absorbance distribution ${\cal P}_A(\alpha)$.


\begin{figure}[h]
\includegraphics[width=1\columnwidth,keepaspectratio,clip]{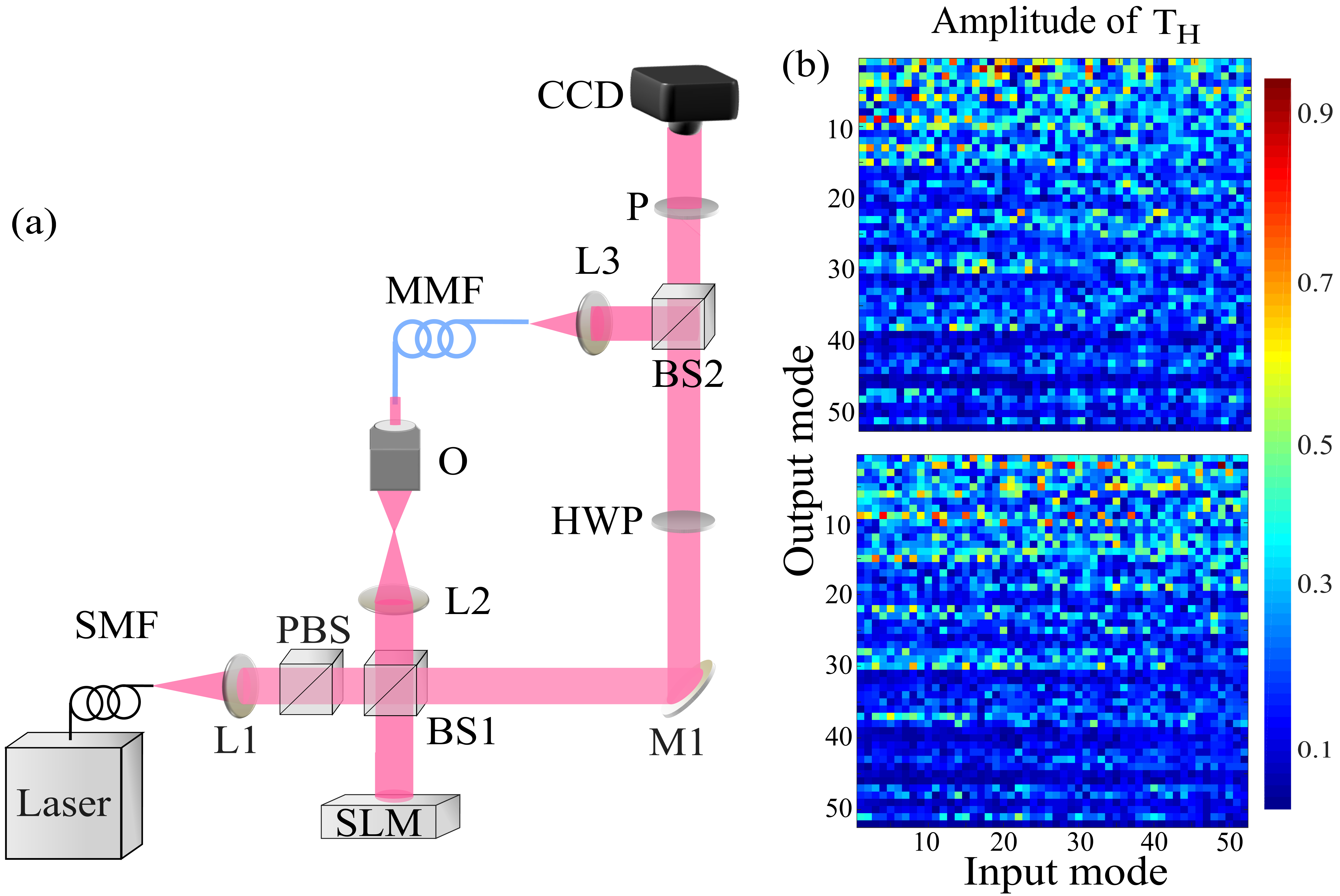}
\caption{ (color online) (a) Schematic of the experimental setup for measuring transmission matrices of a MMF. SMF: single mode fiber. 
L: lens.  PBS: polarizing beam splitter. BS: beam splitter. M: mirror. HWP: half-wave plate. P: polarizer. (b) Amplitude of measured 
transmission matrix for horizontally polarized input light and both horizontally and vertically polarized output fields in the fiber mode 
basis (the top half is $\tilde{t}_{HH}$, and the bottom half is $\tilde{t}_{VH}$)}.
\label{fig2}
\end{figure}

To confirm our theoretical predictions, we experimentally measured the transmission matrices ${\tilde t}_{\rm HH}$ and ${\tilde t}_{\rm 
VH}$ for several realizations of MMFs with strong mode coupling (here and below all measured quantities are indicated with a tilde). 
The polarization-resolved transmission matrix is characterized with an interferometric setup, see Fig. \ref{fig2}. A laser beam at wavelength 
$\lambda = 1550$ nm is collimated by a lens and then horizontally polarized by a polarizing beam splitter. The beam is split into a reference 
arm and a fiber arm.  The SLM in the fiber arm is imaged to the input facet of the fiber by a lens and a microscope objective. It generates 
plane waves with different angles to excite different fiber modes. A half-wave plate rotates the polarization direction of the reference beam. 
Transmitted light from the distal end of the fiber is recombined with light from the reference arm at another beam splitter with a tilt angle, 
forming interference fringes on the CCD camera. A linear polarizer in front of the camera selects the polarization component to measured. 
By rotating the polarizer, we measure the transmitted light of different polarization. The amplitude and phase of the output field are extracted 
from the interference fringes.

\begin{figure}[h]
\includegraphics[width=1\columnwidth,keepaspectratio,clip]{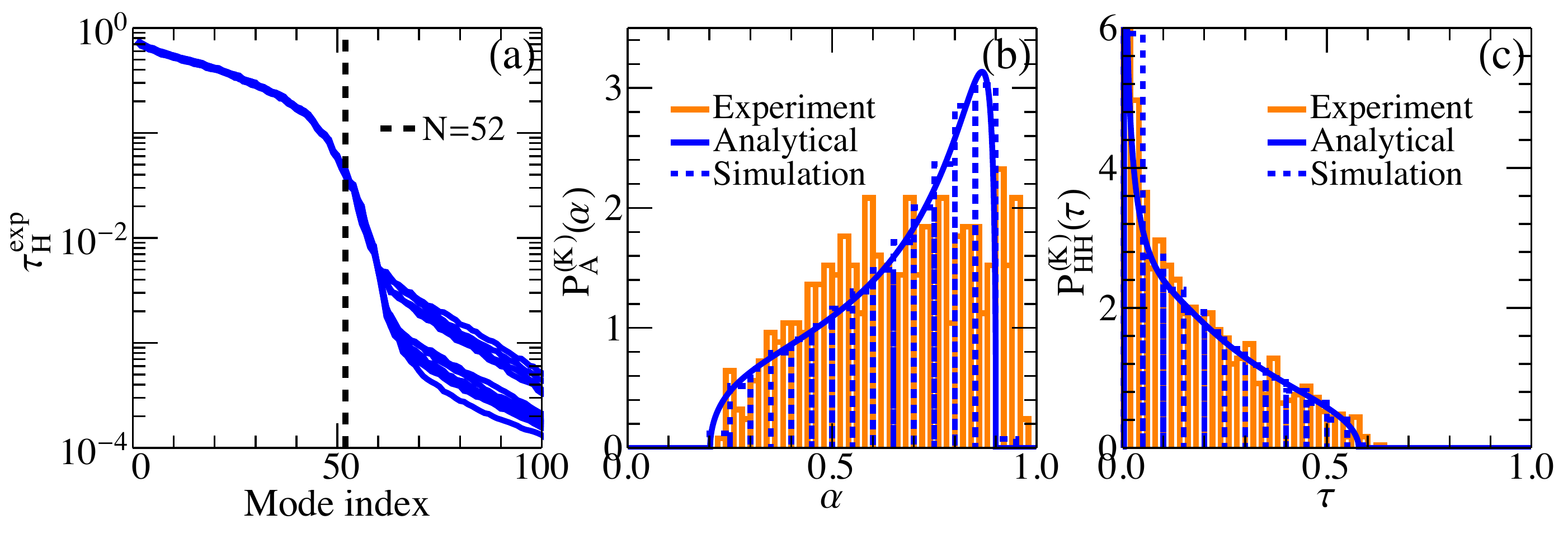}
\caption{ (color online) Experimental results and comparison with theory.
(a) Eigenvalues of the measured matrix ${\tilde T}_{\rm H}$, showing the number of spatial modes to be $N=52$.
(b,c) Measured eigenvalue distributions $\tilde{\cal P}_A^{(K)}(\alpha)$ for the total transmission summed over the two output polarizations and
 $\tilde{\cal P}_{\rm HH}^{(K)}(\tau)$ for the polarization-maintaining output, both showing comparison to the analytic results and to simulations.
}
\label{fig3}
\end{figure}

The MMF we tested is a graded-index fiber with 50 $\mu$m core diameter and 0.22 numerical aperture. To introduce mode mixing in the 
2-meter long bare fiber, we coil the fiber and use clamps to apply stress. The clamps deform the fiber, causing strong mode mixing and 
inevitable mode dependent loss. To determine the number of spatial modes in the fiber, we examine the eigenvalues of the matrix ${\tilde 
T}_{\rm H}=\left({\tilde t}_{\rm HH}^{(K)}\right)^{\dagger}{\tilde t}_{\rm HH}^{(K)}+\left({\tilde t}_{\rm VH}^{(K)}\right)^{\dagger}{\tilde t}_{\rm 
VH}^{(K)}$, as shown in Fig.~\ref{fig3}a. The sudden drop of the eigenvalues corresponds to the cut-off of the guided modes in the fiber 
\cite{Carpenter14}. Base on this analysis, we determine the number of fiber modes to be $N=52$. Subsequently we project the transmission 
matrices ${\tilde t}_{\rm HH}$ and ${\tilde t}_{\rm VH}$ onto the space spanned by the 52 eigenmodes of ${\tilde T}_{\rm H}$ with the largest 
eigenvalues.

To determine the model parameters $K$ and $s$, we evaluate the mean ${\tilde \mu}_H^{(K)}$ and the variance ${\tilde \sigma}_H^{K}$ of 
the eigenvalue distribution $\tilde{\cal P}_H^{(K)}$ of the experimental ${\tilde T}_{\rm H}$ measured over different realizations of the fiber.
A direct comparison with the theoretical predictions Eq.~(S11) yields $K=1$ and $s= 2.7$. In Fig.~\ref{fig3}b we plot the 
theoretical expression Eq. (S10) for the distribution of absorbances ${\cal P}_A(\alpha)$ using the extracted $(K,s)$ parameters (blue 
line). At the same figure we plot the experimental distribution $\tilde{\cal P}_A^{(K)}$ (orange-line histogram) together with results of 
simulations (blue-dashed histogram) from the concatenated MMF model. The agreement is reasonably good.

Finally, we examine the polarization-maintaining eigenvalue distribution $\tilde{\cal P}_{\rm HH}^{(K)}$ evaluated from the experimentally 
measured transmission matrices, and compare it to the analytic prediction of ${\cal P}_{\rm HH}^{(K)}$ from Eqs.~(\ref{eq: HH},\ref{GfuncP}) 
using the extracted $(K,s)$ parameters (Fig.~\ref{fig3}c). We observe excellent quantitative agreement with no fitting, which validates 
our model and our analytic framework.

{\it Conclusions --} We have developed a theoretical formalism that utilizes a free probability theory together with a filtered random matrix approach in order to derive theoretical expressions for the probability distribution of transmittances and absorbances in multimode scattering set-ups where  reflection mechanisms are absent (paraxial approximation) and the information about the transmission matrix is incomplete. The motivation for this study is drawn by the recent interest to understand light transport in MMFs with mode-dependent loss and strong mode and polarization mixing. The resulting probability distributions are different from any known results found for lossy disordered or chaotic systems ~\cite{F03,BB97, B98,SS03} indicating that the paraxial constraint, and/or the presence of MDLs can dramatically affect light transport. The validity of our predictions 
have been tested both with simulations and via direct comparison with experimental data. We stress that our scheme can take into account any type of MDL (or gain) distribution. It will be interesting to extend this study to the case of weak mode mixing where the mode-mixing matrices $v_k$ do not belong to the CUE.

{\it Acknowledgments --}We thank Prof. Stefan Rotter for stimulating discussions. This work is supported partly by NSF under grants No. ECCS-1509361
and No. EFMA-1641109 and AFOSR MURI grant FA9550-14-1-0037 


\begin{center}
\textbf{\large Supplemental Materials}
\end{center}

\section{Derivation of the implicit equation for $G^{\left(K\right)}\left(z\right)$ and distribution of ${\cal P}^{(K)}(\tau)$}

We can construct a recursion relation from the model definition in Eq.~(1) of the main text.
Statistically every segment is equivalent, so we can write
\begin{equation}
\left(t^{(K+1)}\right)^{\dagger}t^{(K+1)} = v_{0}^{\dagger}\Lambda\left(t^{(K)}\right)^{\dagger}t^{(K)}\Lambda v_{0},
\label{eq: recursion}
\end{equation}
where the equality is in the statistical sense.
Using the free probability theory \cite{V87,B13}, when $N\rightarrow\infty$
we get from Eq.~(\ref{eq: recursion})

\begin{equation}
S_{\left(t^{(K+1)}\right)^{\dagger}t^{(K+1)}}\left(z\right)= S_{\left(t^{(K)}\right)^{\dagger}t^{(K)}}\left(z\right)S_{\Lambda^{2}}\left(z\right)
\label{eq: S_recursive}
\end{equation}
where $S_{Q}$ denotes the $S$ transform for an arbitrary hermitian matrix $Q$. The $S$ transform $S_{Q}$ is ultimately related to
the Green's function $G_{Q}\left(z\right)\equiv\int d\lambda\frac{P_{Q}\left(\lambda\right)}{z-\lambda}$, where $P_{Q}\left(\lambda\right)$ 
is the eigenvalue density of the hermitian matrix $Q$. The intermediate connections are shown as follows:
\begin{align}
S_{Q}\left(z\right)= & \frac{z+1}{z}\chi_{Q}\left(z\right),\nonumber \\
\chi_{Q}\left(\phi_{Q}\left(z\right)\right)= & \phi_{Q}\left(\chi_{Q}\left(z\right)\right)=z,\label{eq: connections}\\
\phi_{Q}\left(z\right)= & \frac{1}{z}G_{Q}\left(\frac{1}{z}\right)-1,\nonumber 
\end{align}
where $\phi_{Q}\left(z\right)$ is the moment generating function
and $\chi_{Q}\left(z\right)$ is the corresponding inverse function.
The Green's function $G_{Q}\left(z\right)$ enables us to obtain the
normalized eigenvalue density of the hermitian matrix $Q$ through
the relation
\begin{align}
P_{Q}\left(\lambda\right)= & -\frac{1}{\pi}\lim_{\epsilon\rightarrow0^{+}}\mathrm{Im}G_{Q}\left(\lambda+i\epsilon\right).
\label{eq: Pq}
\end{align}
In the case of one section $K=1$, we have the eigenvalue density
\begin{align}
P_{\left(t^{(1)}\right)^{\dagger}t^{(1)}}\left(\lambda\right)=P_{\Lambda^{2}}\left(\lambda\right)= & \frac{1}{s\lambda}
\label{eq: 1section}
\end{align}
 where $\lambda\in\left(e^{-s},1\right)$ . Correspondingly, we can get from Eq.~(\ref{eq: connections}) the $S$ transform for one
section
\begin{align}
S_{\left(t^{(1)}\right)^{\dagger}t^{(1)}}\left(z\right)= & S_{\Lambda^{2}}\left(z\right)=\frac{z+1}{z}\frac{e^{sz}-1}{e^{sz}-e^{-s}}.
\label{eq: S_1sec}
\end{align}
Combining Eq.~(\ref{eq: S_recursive}) and Eq.~(\ref{eq: S_1sec}) we have 
\begin{align}
S_{\left(t^{(K)}\right)^{\dagger}t^{(K)}}\left(z\right) & =\left(\frac{z+1}{z}\frac{e^{sz}-1}{e^{sz}-e^{-s}}\right)^{K}.
\label{eq: S_K}
\end{align}
Thus we can use Eq.~(\ref{eq: connections}) and Eq.~(\ref{eq: S_K})
to get the implicit formula for $G^{\left(K\right)}\left(z\right)$,
which is shown in Eq. (2) of the main text. 

At this stage typically we resort to numerical method to obtain the eigenvalue density by combing Eq.~(\ref{eq: Pq}) and the implicit 
formula for the Green's function, say Eq.~(2) corresponding to the $K$-section full transmission matrix. However, for the 
first few moments, explicit results can be easily obtained. For example, using Eq.~(\ref{eq: S_recursive}) and (\ref{eq: connections})
we can get the mean $\mu^{(K)}$ and variance $(\sigma^{(K)})^{2}$ for the eigenvalue density of $\left(t^{(K)}\right)^{\dagger}t^{(K)}$ as
\begin{align}
\mu^{(K)}= & \mu^{K},\:\left(\frac{\sigma^{(K)}}{\mu^{(K)}}\right)^{2}=K\frac{\sigma^{2}}{\mu^{2}},
\end{align}
where $\mu$ and $\sigma^{2}$ are the mean and the variance of the
eigenvalue distribution Eq.~(\ref{eq: 1section}) for one section,
which turn out to be 
\begin{align}
\mu= & \frac{1-e^{-s}}{s},\\
\sigma^{2}= & \frac{1-e^{-2s}}{2s}-\left(\frac{1-e^{-s}}{s}\right)^{2}.\nonumber 
\end{align}

\begin{figure}[h]
\includegraphics[width=1\columnwidth,keepaspectratio,clip]{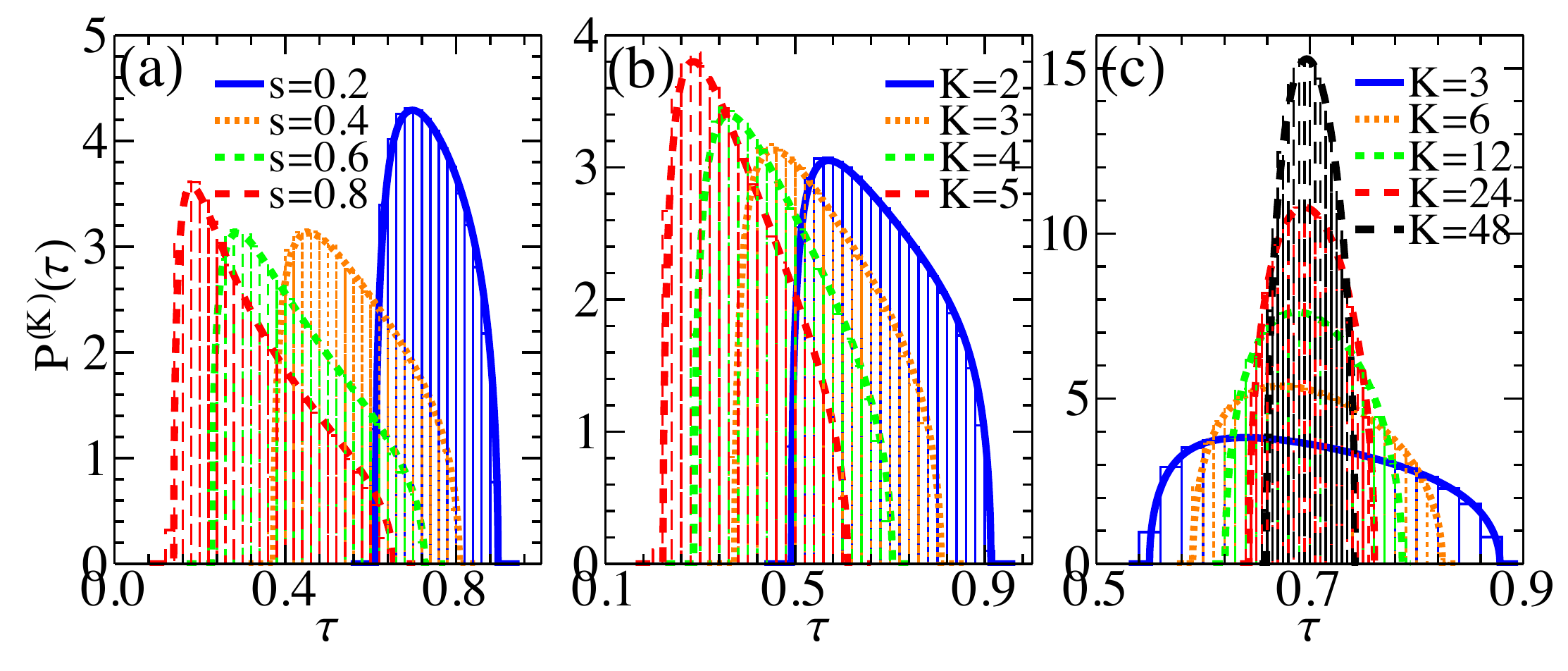}
\caption{ (color online) Distribution ${\cal P}^{(K)}(\tau)$ for (a) different $s$-values and fixed $K=3$; (b) different $K$-values 
and fixed $s=0.4$. In both cases the number of modes is $N=25$. (c) $K\rightarrow \infty$, $s\rightarrow 0$ while $\mu^{(K)}=0.7$ is 
kept fixed. In this case $N=75$.}
\label{Sfig1}
\end{figure}

Finally using the relation Eq. (\ref{eq: Pq}) we can calculate the probability distribution ${\cal P}^{(K)}(\tau)$. In Fig. \ref{Sfig1} we show the 
theoretical results together with the outcome of the simulations. We find that for finite number of concatenated segments $K$ and finite MDL
$s\ne 0$, the distribution ${\cal P}^{(K)}(\tau)$ deviates from the standard semicircle expected from standard RMT considerations, see Figs. 
\ref{Sfig1}a,b. The latter shape is restored (around the mean $\mu^{(K)}$) in the limit $K\rightarrow \infty$, $s\rightarrow 0$ while $\mu^{(K)}$ 
is kept constant, see Fig. \ref{Sfig1}c.

\section{Distribution of Absorbance eigenvalues}

Following exactly the same steps we can also evaluate, using the FRM formalism, the Green's function $G_H^{(K)}$ 
associated with the total (irrespective of the output polarization state) transmittance matrix $T\equiv \left(t_{\rm HH}^{(K)}
\right)^{\dagger}t_{\rm HH}^{(K)}+\left(t_{\rm VH}^{(K)}\right)^{\dagger}t_{\rm VH}^{(K)}$ of an incident light prepared in the H-
polarization. The complementary matrix $A\equiv 1- T$ indicates the relative absorption occurring during its propagation 
inside the MMF. 

In case of lossless fibers, i.e. $s=0$, the total transmittance is unity. When $s
\ne 0$ we get a similar relation for $G_{\rm H}^{(K)}$ and $G^{(K)}$ as Eq. (3) where now $d_{\rm HH}\rightarrow 
d_{\rm H}={1\over 4}\left(z G_{\rm H}^{(K)}+1\right)G_H^{(K)}$ while $n_{\rm H}=n_{\rm HH}$. Using the modified Eq. (3) 
together with Eq. (2) we get the following implicit formula for $G_{\rm H}^{K}$
\begin{align}
\frac{1}{z}= & \left(\frac{p_H}{p_H-1}\right)\left(\frac{p_H}{p_H-2}\right)^{K-1}\left(\frac{e^{\frac{s}{2}p_H}
-e^{s}}{e^{\frac{s}{2}p_H}-1}\right)^{K}
\label{eq: H}
\end{align}
where $p_H\left(z\right)=zG_H^{(K)}\left(z\right)+1$. The distribution ${\cal P}_{\rm H}^{(K)}(\tau)$ associated with the 
eigenvalues of $T$-matrix is then given via Eq. (5) by substituting $G_{\rm HH}^{(K)}\rightarrow G_{\rm H}^{(K)}$
while the corresponding mean value $\mu_{\rm H}^{(K)}$ and variance $\left(\sigma_{\rm H}^{(K)}\right)^{2}$ can be expressed
in terms of the microscopic variables of the concatenated model as
\begin{align}
\mu_{\rm H}^{(K)}= \mu^{(K)}=\mu^{K},\:\left(\frac{\sigma_{\rm H}^{(K)}}{\mu_{\rm H}^{(K)}}\right)^2=\frac{1}{2}\left(\frac{\sigma^{(K)}}
{\mu^{(K)}}\right)^2={K\over 2}\left({\sigma \over \mu}\right)^2
\label{eq: mean_H}
\end{align}

\section{Analysis of experimental data }

Experimentally, for the finite MMF, we get an ensemble of the unscaled
quarter $t_{\rm HH}$ and half $t_{\rm H}\equiv\begin{pmatrix}t_{\rm HH}\\
t_{\rm VH}
\end{pmatrix}$ of the transmission matrix $t$ in the channel space, where both
$t_{\rm HH}$ and $t_{\rm VH}$ are of dimension $14641\times441$ ($121\times 121$ camera pixels and $21\times 21$ input angles). The number
of relevant modes in the fiber is estimated to be $N=52$ (see the
main text). We use the specific experimental measurement to normalize
the unscaled date of the transmission matrix $t_{\rm H}$. Experimentally
we excite the low-order modes through the single-channel input $e_{1}=\left(\begin{array}{cccc}
1 & 0 & \cdots & 0\end{array}\right)^{T}$. The low-order-mode space is specified as follows. Consider the eigendecomposition
\begin{align}
t_{\rm H}^{\dagger}t_{\rm H}= & w\Lambda_{d}w^{\dagger}
\end{align}
where $\Lambda_{d}$ is a real diagonal matrix with the diagonal entries
sorted in decreasing order and $w$ is the unitary matrix with columns
being the eigenvectors of $t_{\rm H}^{\dagger}t_{\rm H}$. Let $\tilde{w}$
be the truncation of the matrix $w$ obtained from taking the first
$N=52$ columns. Thus the low-order-mode space is spanned by the columns
of $\tilde{w}$. The excitation of the single-channel input $e_{1}$
in this low-order-mode space is given as 

\begin{align}
v= & \tilde{w}\tilde{w}^{\dagger}e_{1}.
\end{align}
Experimentally we have
\begin{align}
\frac{v^{\dagger}\tilde{t}_{\rm H}^{\dagger}\tilde{t}_{\rm H}v}{v^{\dagger}v}= & 0.48,\label{eq: exp_constraint}
\end{align}
where $\tilde{t}_{\rm H}=ct_{\rm H}$ is the normalized data for the transmission
matrix. Eq.~(\ref{eq: exp_constraint}) enables us to determine the
scaling constant $c$ and thus properly normalized transmission data
$\tilde{t}_{\rm H}$. Correspondingly we can extract from $\tilde{t}_{\rm H}$
the half truncation matrix $\tilde{t}_{\rm HH}$ and $\tilde{t}_{\rm VH}$. 



\end{document}